\documentstyle[preprint,aps]{revtex}
\begin{document}
\draft
\title{\bf Failure of hydrodynamics within the vortex liquid phase}

\author{T. Blum and M.A. Moore}

\address{Department of Theoretical Physics}
\address{University of Manchester}
\address{Manchester M13 9PL}
\address{United Kingdom}

\date{\today}
\maketitle

\begin{abstract}
The recent discovery that some of the coefficients of the viscosity
tensor are negative is shown to invalidate the hydrodynamic approach
to the vortex liquid phase of a type-II superconductor.
\ A satisfactory theory requires retention of all the spatial gradients
of the velocities or electric fields and not just the first derivatives,
as assumed in a hydrodynamic theory.
\ We illustrate such a procedure by using time-dependent Ginzburg-Landau
theory to determine the electric field distribution near a single
``twin-plane boundary" due to a current passing through the boundary.
\
\end{abstract}

\vfill


\newpage
\narrowtext

There exists an extensively developed phenomenological hydrodynamic
theory \cite{MarNel1,MarNel2,Mar1,HusMaj,RadFrey,ChenMar} for
superconductors in the vortex liquid regime.
\ In the recent work of Mou {\it et al.} \cite{Mouetal}, the coefficients
of the viscosity tensor were calculated from a more fundamental theory,
in this case Ginzburg-Landau theory, and some were found to be negative.
\ We show explicitly that this invalidates the hydrodynamic approach.
\ We find that a satisfactory theory requires retention of all the spatial
gradients of the velocities and not just the first derivatives, as
assumed in a hydrodynamic theory.
\ We illustrate our procedure by calculating the electric field
distribution near a single ``twin-plane boundary" when a constant
current is being passed through that boundary.
\ The electric field decays to its bulk value; whereas, the hydrodynamic
theory with negative viscosity coefficients (as calculated) has
unphysical oscillatory behavior of the electric field as a function of
distance from the boundary.

Because the motion of extended flux lines plays a vital role, the transport
properties of type-II superconductors near and below the $H_{c2}(T)$ line
are expected to be nonlocal.
\ For instance, the current ${\bf j}$ produced by a small electric
field ${\bf E}$, is given in terms of the conductivity tensor, as:
\begin{equation}
j_{\mu}({\bf r}) \ = \ \int \sigma_{\mu \nu}({\bf r},{\bf r^{\prime}})
{}~E_{\nu}({\bf r^{\prime}}) ~d{\bf r^{\prime}}
\end{equation}
where $\sigma_{\mu\nu}$ is nonlocal, {\it i.e.}
$\sigma_{\mu \nu}({\bf r},{\bf r^{\prime}}) \neq \sigma_{\mu \nu}
\delta({\bf r}-{\bf r^{\prime}})$.
\ Assuming translational invariance, one has then in momentum space:
\begin{equation}
j_{\mu}({\bf k}) \ = \ \sigma_{\mu \nu}({\bf k}) ~E_{\nu}({\bf k}),
\end{equation}
where nonlocality now implies $\sigma_{\mu \nu}({\bf k}) \neq Const.$
\ The nonlocality of the conductivity has been recently investigated
experimentally. \cite{Safar}

The ``hydrodynamic" approach
\cite{MarNel1,MarNel2,Mar1,HusMaj,RadFrey,ChenMar}
assumes that an adequate description of the long-time,
large-distance behavior can be derived from a truncation of the
small-wave-vector, small-frequency
expansion of the conductivity to the lowest non-trivial terms.
\ For instance, the dc hydrodynamic conductivity tensor would be
taken in the hydrodynamic approach to be:
\begin{equation}
\sigma_{\mu \nu}({\bf k},\omega=0) \ = \ \sigma_{\mu \nu}(0)
\ + \ S_{\mu \alpha \beta \nu}~k_{\alpha}k_{\beta},
\label{hydro}
\end{equation}
(in the notation of Mou {\it et al.}\cite{Mouetal}).
\ Moreover, hydrodynamics includes a tacit assumption that this truncation
results in a stable theory; $S_{zzzz}$, for example, must be positive.
\ (We take the external field ${\bf B}$ to be in the $\hat z$ direction.)

However, recent calculations of Mou {\it et al.} \cite{Mouetal}
of the $S$'s from Ginzburg-Landau theory found some of them
to have signs opposite of those required for stability.
\ We extend their calculations and show that stability is regained
if the full expansion in powers of the wave vector is considered.
\ Then we proceed to calculate the ${zz}$ component of the
conductivity in the presence of a boundary and with it, we calculate
the electric field for the case of a constant current $j_z$ being
passed through a ``twin-plane boundary" (at $z=0$).
\ The results are compared with those from the hydrodynamic theory.
\cite{MarNel2}
\ This example also illustrates the necessity for stability of including
the normal component of the conductivity (assumed to be local).

Our starting point is the time-dependent Ginzburg-Landau equation:
\begin{equation}
\left[{1 \over \Gamma} {\partial \over \partial t} -
\sum_{j=1}^3 {\hbar^2 \over 2m_j} \left( {\partial \over
\partial r_j} -{ie^*A_j \over \hbar} \right)^2 +a
\right] \psi({\bf r},t)
{}~+~{b \over 2}~|\psi({\bf r},t)|^2~\psi({\bf r},t)
\ = \ \eta({\bf r},t),
\label{Landau}
\end{equation}
with noise correlations:
\begin{equation}
\left\langle \eta^*({\bf r},t)~\eta({\bf r^{\prime}},t^{\prime})
\right\rangle \ = \ {2 k_BT \over \Gamma}
{}~\delta({\bf r}-{\bf r^{\prime}}) ~\delta(t - t^{\prime})
\label{noise}
\end{equation}
and with $m_{1,2}=m_{ab}$ and $m_3=m_c$.
\ We use the symmetric gauge
\begin{equation}
A_x \ = \ {-By \over 2} \ \ \ \ {\rm and} \ \ \ \
A_y \ = \ {Bx \over 2} .
\end{equation}
\ As the parameter $\Gamma$ sets the time scale, time will here after
be measured in units of $\Gamma$; $\hbar$ will be set to one.
\ The following calculations are based on the linearized version of eq.
(\ref{Landau}), and so strictly speaking above the $H_{c2}(T)$ line;
however, making the Hartree approximation:
\begin{equation}
a_{eff} \ = \ a ~+~ b~\langle |\psi|^2 \rangle
\end{equation}
can take one below the $H_{c2}(T)$ line. \cite{RugTho}

First, let us sketch out the derivation of the bulk conductivity,
returning to the case with a boundary afterward.
\ The linear response to the field can be calculated via the Kubo
formula which relates the conductivity to the current fluctuations
as follows:
\begin{equation}
\sigma^{(s)}_{\mu \nu}({\bf k},\omega) \ = \ {1 \over 2 k_B T}
\int d({\bf r}-{\bf r^{\prime}}) \int d(t-t^{\prime})
{}~{\rm e}^{i{\rm k}\cdot ({\rm r}-{\rm r^{\prime}})
-i \omega (t - t^{\prime}) }
{}~\langle J_{\mu}({\bf r},t)~J_{\nu}({\bf r^{\prime}}, t^{\prime})
\rangle_c~.
\label{kubo1}
\end{equation}
Here, we will be interested in the non-local nature of the dc
conductivity: $\sigma^{(s)}_{\mu\nu}({\bf k},\omega=0)$.
\ (The frequency dependence of the uniform conductivity
$\sigma^{(s)}_{\mu\nu}({\bf k}={\bf 0},\omega)$ was studied
in ref. \cite{Schmidt}.)

In order to calculate the current fluctuations, we will need the
Green's function and correlation function.
\ Instead of the usual eigenfunction expansion, we take the
result for the Green's function from the path-integral approach
\cite{Feynman}:
\begin{eqnarray}
 G({\bf r}, t^{\prime}+\tau; {\bf r^{\prime}},t^{\prime}) &\ = \ &
\left({m_c \over 32 \pi^3 \ell^4
\tau}\right)^{1/2}
{}~{ 1 \over {\rm sinh}(\omega_o \tau/2 ) }
\ {\rm exp}\left\{-a \tau -{m_c \over 2 \tau}
(z-z^{\prime})^2 \right\}
\nonumber \\
&&{\rm exp} \left\{~-{{\rm coth}(\omega_0\tau/2) \over 4 \ell^2}
\left[(x-x^{\prime})^2 + (y-y^{\prime})^2 \right]
{}~+~{i \over 2 \ell^2} (xy^{\prime}-yx^{\prime})~\right\},
\label{Green}
\end{eqnarray}
where $\omega_o=e^*B /m_{ab}$ is the cyclotron frequency,
$\ell=(e^*B)^{-1/2}$ is the magnetic length, and $\tau > 0$.

{}From the definition of a Green's function:
\begin{equation}
\psi({\bf r},t) \ = \ \int dt^{\prime} \int d {\bf r^{\prime}} ~
G({\bf r},t;{\bf r^{\prime}}, t^{\prime})
{}~\eta({\bf r^{\prime}}, t^{\prime}),
\end{equation}
which one can then use to calculate the correlation function
$\left\langle \psi^*({\bf r^{\prime}},t^{\prime}) ~\psi({\bf r}, t)
\right\rangle$.
\ Using eq. (\ref{noise}) for the noise correlations, it follows
after some algebra that:
\begin{equation}
\left\langle \psi^*({\bf r^{\prime}},t^{\prime}) ~\psi({\bf r}, t)
\right\rangle \ = \ 2 k_B T  \int_0^{\infty}
ds ~ G({\bf r},{\bf r^{\prime}};|t-t^{\prime}|+2s).
\label{correlator}
\end{equation}

The Kubo formula for the conductivity involves a product of correlation
functions with some (covariant) derivatives acting on it, those
derivatives coming from the usual definition of the current density:
\begin{equation}
J_{\mu}({\bf r},t) \ = \ {e^* \over 2 m_{\mu}} ~\left[
\psi^* \left(-i {\partial \over \partial r_{\mu}}
-e^*A_{\mu} \right)\psi ~+~
\psi ~\left(i {\partial \over \partial r_{\mu}}
-e^*A_{\mu} \right) \psi^* \right].
\label{current}
\end{equation}
The next steps of the calculation are relatively straightforward:
\ (i) insert eqs. (\ref{Green}), (\ref{correlator}) and (\ref{current})
into the Kubo formula (\ref{kubo1}); (ii) carry out the required
derivatives; and (iii) perform the spatial integrations in eq.
(\ref{kubo1}), which are all Gaussian integrals.
\ Three integrations over time variables (one from the Kubo formula
and one from each correlation function) remain.
\ After some rearrangement of these time integrations, we find:
\begin{eqnarray}
\sigma^{(s)}_{\mu \nu}({\bf k},0) &\ = \ &{e^{*2} k_BT m_c^{1/2}\over
8 \pi^{3/2}m_{\mu}m_{\nu} \ell^2}
\int_0^{\infty} d\tau ~{\tau^{3/2}~{\rm e}^{-2a\tau} \over
{\rm sinh}(\omega_0 \tau)} \int_0^1 dv ~v \int_{0}^{1} du
 ~C_{\mu \nu}(\tau,u,v,{\bf k}) \nonumber \\
&&{\rm exp} \left\{~ -~{ \ell^2 [{\rm cosh}(\omega_0 \tau) -
{\rm cosh}(uv \omega_0 \tau)] \over 2
{}~{\rm sinh}(\omega_0 \tau) }
{}~(k_x^2+k_y^2) ~-~{\tau (1-u^2v^2) \over 4 m_c} ~k_z^2
{}~\right\},
\label{bulk1}
\end{eqnarray}
where
\begin{eqnarray}
&&C_{xx}\ =
{2~{\rm cosh}(uv\omega_0 \tau) \over
\ell^2~{\rm sinh}(\omega_0 \tau)}
{}~+~ { {\rm sinh}^2(uv\omega_0 \tau)
 \over {\rm sinh}^2(\omega_0 \tau)} ~k_{x}^2
{}~+~ { [{\rm cosh}(\omega_0 \tau)~-~
{\rm cosh}(uv\omega_0 \tau)]^2
 \over {\rm sinh}^2(\omega_0\tau)} ~k_{y}^2 \nonumber \\
&&C_{zz} \ = \  {2 m_c \over \tau} ~+~ u^2v^2 ~k_z^2 \nonumber \\
&&C_{xy}\ =\
{ {\rm sinh}^2(uv\omega_0 \tau)~-~
[{\rm cosh}(\omega_0\tau)~-~
{\rm cosh}(uv\omega_0\tau)]^2
 \over {\rm sinh}^2(\omega_0\tau)} ~k_xk_y \nonumber \\
&&C_{xz}\ = \
{uv~{\rm sinh}(uv\omega_0 \tau) \over
{\rm sinh}(\omega_0\tau)} ~k_xk_z.
\label{bulk2}
\end{eqnarray}
The $\{x \leftrightarrow y\}$ symmetry can be used to obtain the
other $C_{\mu\nu}$'s.

Expanding eq. (\ref{bulk1}) to quadratic order in ${\bf k}$ and
performing the $u$ and $v$ integrations reproduces the results of
Mou {\it et al.} \cite{Mouetal}, including results such as $S_{xxxx}<0$
and $S_{zzzz}<0$.
\ However, one can see from eqs. (\ref{bulk1}) and (\ref{bulk2}) that
the full expression for $\sigma^{(s)}_{ii}({\bf k})$ is greater than zero
as required for stability.

In some special instances, we can resum the expansion in powers of
${\bf k}$; a case we will be using below is
$\sigma^{(s)}_{zz}(k_x=0,k_y=0,k_z)$:
\begin{equation}
\sigma^{(s)}_{zz}(k_z)\ =\
{e^{*2} m_c^{1/2}k_BT \over \pi \ell^2 }
{}~{1 \over k_z^2} ~\sum_{n=0}^{\infty} \left[
\left(\alpha +2 n \omega_0 \right)^{-1/2} -
\left(\alpha + 2 n \omega_0 +k_z^2/4m_c\right)^{-1/2}
\right],
\label{zzcomp1}
\end{equation}
where
\begin{equation}
\alpha\ = \ 2a ~+~ \omega_0
\end{equation}
measures the distance from the $H_{c2}(T)$ line.

The above results are, however, for the bulk conductivity.
\ In the case with boundaries, the calculation should be repeated
with the Green's function ${\cal G}$ which satisfies the boundary
conditions imposed on the Ginzburg-Landau equation (\ref{Landau}).
\ Let us consider an example with a boundary condition at $z=0$:
\begin{equation}
\psi (x,y,z=0;t) \ = \ 0.
\end{equation}
One could imagine that such a boundary condition might arise from
having a defect like a twin-plane at $z=0$.
\ We admit that this is not a realistic model in any sense for
such a defect.
\ It is just the simplest example we could devise which enables
us to make some analytical progress.
\ The response function ${\cal G}$ is then given by:
\begin{equation}
{\cal G}(x,y,z;x^{\prime},y^{\prime},z^{\prime};t)
\ = \ G(x,y,z;x^{\prime},y^{\prime},z^{\prime};t)
{}~-~ G(x,y,-z;x^{\prime},y^{\prime},z^{\prime};t),
\end{equation}
which is just the difference between the bulk Green's function and
its image.

Suppose there is a constant current in the $\hat z$ direction being
passed through this boundary, and that one would like to calculate
the corresponding electric field.
\ Since in this situation the current and electric field both point in the
$\hat z$ direction, only the $zz$ component of the conductivity
$\Sigma^{(s)}$ in the presence of the boundary is required.
\ Furthermore, the electric field varies only in the $\hat z$ direction,
hence, the $x$ and $y$ variables can be summed over (or in momentum
space $k_x=0$, $k_y=0$).
\ Thus we need to evaluate $\Sigma^{(s)}_{zz}(z,z^{\prime})$.
\ As from now on we shall only be considering $z$ components, we shall
drop the $z$ indices; for example, $E_z \rightarrow E$,
$\sigma_{zz}^{(s)}({\bf k}) \rightarrow \sigma^{(s)}({\bf k})$,
$S_{zzzz} \rightarrow S$, etc.
\ Furthermore, we will denote the uniform conductivity $\sigma^{(s)}(0)$
simply as $\sigma^{(s)}$.

The calculation of the conductivity (which is a four-point function)
involves a product of Green's functions.
\ Consequently, when repeating the steps outlined earlier with ${\cal G}$
instead of $G$, one can identify four pieces:
\begin{equation}
\Sigma^{(s)}(z,z^{\prime}) \ = \
\sigma^{(s)}(z-z^{\prime}) ~-~
\sigma^{(s)}(z+z^{\prime}) ~+~
\varsigma^{(s)}(z,z^{\prime}) ~-~
\varsigma^{(s)}(-z,z^{\prime}).
\label{sigma}
\end{equation}
The first term comes from the product $G(z-z^{\prime})G(z-z^{\prime})$
and is, of course, the bulk conductivity $\sigma^{(s)}(z-z^{\prime})$,
the Fourier transform of which was calculated above, eq. (\ref{zzcomp1}).
\ The second term arises from $G(z+z^{\prime})G(z+z^{\prime})$ and
is the image of the bulk conductivity $-\sigma^{(s)}(z+z^{\prime})$; the
negative sign resulting from the derivative with respect to $z$ associated
with the current operator.
\ The third term originates from the cross term $G(z-z^{\prime})
G(z+z^{\prime})$ and takes on the form:
\begin{eqnarray}
\varsigma(z,z^{\prime})\ &=& \
{e^{*2} k_BT \over 32 \pi^2 m_c \ell^2}
\int_0^{\infty} d\tau ~\tau \int_0^1 dv ~v \int_{-1}^1du
{}~{{\rm e}^{-2a\tau} \over {\rm sinh}(\omega_0 \tau)
{}~(1-u^2v^2)^{1/2}}
\nonumber \\
&&\times \left[ {2 m_c uv \over \tau (1-u^2v^2) }
{}~-~{m_c^2 \over \tau^2 (1+uv)^2}(z-z^{\prime})^2
{}~+~{m_c^2 \over \tau^2 (1-uv)^2}(z+z^{\prime})^2
\right]
\nonumber \\
&&\times {\rm exp} \left\{
{}~-~{m_c \over 2 \tau (1+uv)} (z-z^{\prime})^2
{}~-~{m_c \over 2 \tau (1-uv)} (z+z^{\prime})^2
{}~\right\}.
\end{eqnarray}
And finally, the fourth term is the image of the third, again with a
negative sign.

The first piece, the bulk conductivity $\sigma^{(s)}(z-z^{\prime})$,
decays away from the source at $z^{\prime}$; the second piece, its
image, decays away from the image of the source at $-z^{\prime}$.
\ The cross term $\varsigma(z,z^{\prime})$, on the other hand,
decays away from both source and image.
\ As a consequence, it could only produce a significant contribution
near the boundary where it can be close to both source and image.
\ However, at the boundary, its contribution is killed off by its image
$-\varsigma(-z,z^{\prime})$.
\ Thus, one expects the cross terms to play a lesser role.
\ For this reason, though mainly for reasons of calculational convenience,
we will drop the cross terms.
\ (Of course, the cross terms while cumbersome could be handled
in numerical work if quantitative results were required.)

Thus far the calculation has only concerned the superconducting
contribution to the conductivity.
\ Before proceeding to determine the electric field, the normal
component must be included.
\ Here we will simply add a normal contribution $\sigma^{(n)}$,
neglecting any interference between the two.
\ On the scales of interest, the normal conductivity is local.
\ Hence our final expression for the $zz$ component of combined
conductivity is:
\begin{equation}
\Sigma(z,z^{\prime})\ = \
{1 \over 2\pi} \int_{-\infty}^{\infty}dk_z~
{\rm e}^{-ik_zz^{\prime}}
\left({\rm e}^{ik_zz}-{\rm e}^{-ik_zz} \right)
\left[\sigma^{(n)}+\sigma^{(s)}(k_z) \right].
\label{total}
\end{equation}
Since in the present scenario, we know the current and want the
electric field, we need the inverse of $\Sigma(z,z^{\prime})$,
the resistivity $\rho(z,z^{\prime})$.
\ For an operator of the form (\ref{total}), the inversion is readily
performed:
\begin{equation}
\rho(z,z^{\prime})\ = \
{1 \over 2\pi} \int_{-\infty}^{\infty}dk_z~
{\rm e}^{-ik_zz^{\prime}}
\left({\rm e}^{ik_zz}-{\rm e}^{-ik_zz} \right)
\left[\sigma^{(n)}+\sigma^{(s)}(k_z) \right]^{-1}
\end{equation}
(this ease of inversion is the reason why the ``cross terms" in eq.
(\ref{sigma}) were dropped.)
\ Notice that without the inclusion of $\sigma^{(n)}$, the normal
component, this integral would have been divergent.
\ The electric field is then obtained by integrating over the
source $j(z^{\prime})$:
\begin{equation}
E(z) \ = \ \int_0^{\infty} dz^{\prime}~
\rho(z,z^{\prime}) ~j(z^{\prime}).
\end{equation}
For a constant current $j(z^{\prime})=j_0$, one finds:
\begin{equation}
E(z) \ = \ {2~j_0 \over \pi}~\int_0^{\infty} dk_z~
{{\rm sin}(k_zz) \over k_z
\left[\sigma^{(n)}+\sigma^{(s)}(k_z) \right]}.
\label{electric}
\end{equation}

Inserting the form of $\sigma^{(s)}(k_z)$ provided in
eq. (\ref{zzcomp1}) provides the electric field distribution
shown in Figure 1.
\ At the boundary the electric field $E_z$ takes on the value
$j_0/\sigma^{(n)}$.
\ Since we imposed the boundary condition $\psi(z=0)=0$, the
current there must be purely normal in composition.
\ The electric field then decays toward its bulk value
$j_0/[\sigma^{(n)}+\sigma^{(s)}]$.
\ Within the bulk, the superconducting channel becomes available
and thus a smaller electric field is required to produce the same
current.
\ In expression (\ref{electric}), the length scale associated with
the decay of the field to its bulk value depends not only on static
quantities such as $\alpha$ (which is related to the $c$-axis correlation
length $\xi \sim (m_c\alpha)^{-1/2}/2$) but also on dynamic quantites
such as the normal conductivity $\sigma^{(n)}$.

What would have resulted from the hydrodynamic approach?
\ With a negative coefficient $S$ (as calculated), eq.
(\ref{electric}) would give:
\begin{equation}
E(z) \ = \ {j_0 \over \sigma^{(n)} +\sigma^{(s)}}
\left\{ 1-{\rm cos}\left[\left({\sigma^{(n)}+\sigma^{(s)}
\over |S| }\right)^{1/2}z\right]\right\},
\label{el-hydro}
\end{equation}
yielding unphysical oscillatory behavior (see Figure 1).
\ Even assuming $S>0$ as required for a stable hydrodynamic
theory \cite{MarNel2,HusMaj}, eq. (\ref{electric}) would yield:
\begin{equation}
E(z) \ = \ {j_0 \over \sigma^{(n)} +\sigma^{(s)}}
\left\{ 1-{\rm exp}\left[-\left({\sigma^{(n)}+\sigma^{(s)}
\over S }\right)^{1/2}z\right]\right\},
\label{el-hydro2}
\end{equation}
which increases from an initial value of zero at the boundary and
saturates at the bulk value, in direct contrast to the result
above where the field was largest at the boundary.
\ Another fix is required; the appropriate boundary condition
($E(0)=j_0/\sigma^{(n)}$) must be applied.
\ The correct phenomenology can be achieved in the hydrodynamic method
by using the correct boundary condition and assuming $S$ to be
positive.
\ However, with the latter step, one forfeits making contact with
some underlying theory, which must be seen as a failure of this
approach.

The attraction of the hydrodynamic approach is its simplicity.
\ The calculation considered above is cumbersome even for a single
boundary and with simplifying assumptions.
\ The complications will only multiply as one considers geometries
more appropriate for modeling multiterminal transport measurements.
\ We suggest the following ad hoc procedure which has the virtue of
simplicity, gives reasonable agreement with the full theory, and
which could be generalized to other geometries.
\ We approximate the conductivity by:
\begin{equation}
\sigma(k_z) \ = \ {\sigma^{(n)}~\sigma^{(s)}+
[\sigma^{(s)}]^2+\sigma^{(n)}~|S|~k_z^2 \over
\sigma^{(s)}+ |S|~k_z^2},
\label{simplesig}
\end{equation}
which has the correct small-$k$ behavior ($\sigma^{(n)}+
\sigma^{(s)}+S~k_z^2$, as derived from Landau-Ginzburg theory),
has the correct large-$k$ behavior (\ $\sigma^{(n)}$), and is
positive definite ({\it i.e.} stable).
\ With this form for $\sigma(k_z)$, eq. (\ref{electric}) yields:
\begin{equation}
E_z(z) \ = \ {j_0 \over \sigma^{(n)} + \sigma^{(s)}}
\left\{ 1 +
{\sigma^{(s)} \over \sigma^{(n)} }
{\rm exp} \left[ -\left(
{\sigma^{(s)} \left[\sigma^{(n)} + \sigma^{(s)} \right]
\over \sigma^{(n)} ~|S| }
\right)^{1/2} z \right]
\right\},
\label{simpleE}
\end{equation}
which has the correct limits: \ $E(0)=j_0/\sigma^{(n)}$ at the
boundary and $E(\infty)=j_0/(\sigma^{(n)}+\sigma^{(s)})$
deep inside the bulk.
\ Furthermore, it provides a simple expression for the dependence
of the length scale associated with the decay of $E$ to its bulk
value.
\ Its similarity to the full expression can be judged from Figure
1.

As multiterminal transport measurements represent one of the key
experimental probes of superconducting materials, calculations within
such geometries are important for relating theory and experiment.
\ Steps toward this end were initiated within the phenomenological
hydrodynamic approach \cite{HusMaj}; however, the subsequent calculation
of negative viscosity coefficients \cite{Mouetal} within Ginzburg-Landau
theory invalidates the hydrodynamic approach.
\ Here, we have presented a computation which while more cumbersome than
hydrodynamics has the advantage that it begins with a more
fundamental theory and leads to sensible results.

\

M.A.M. thanks A.T. Dorsey for introducing him to the problems posed
by negative viscosity coefficients.
\ T.B. thanks A.J. Bray for useful discussions.


\bibliographystyle{unsrt}

\begin{figure}
\end{figure}

\begin{description}

\item{Fig. 1}
The solid line shows the electric field (normalized by $j_0/\sigma^{(n)}$)
as a function of distance from the boundary (measured in units of
$(m_c \alpha)^{-1/2}/2$, the mean-field c-axis correlation length) as
calculated from eq. (\ref{electric}) with $\sigma^{(s)}(k_z)$ given by eq.
(\ref{zzcomp1}) with $\sigma^{(n)}/\sigma^{(s)}=1$ and
$\omega_0/\alpha=2$.
\ The dashed line shows $E(z)$ resulting from eq. (\ref{simpleE}).
\ The third line shows is the oscillatory ``hydrodynamic" result using
the same parameters with $S<0$ (as calculated).

\end{description}

\end{document}